\documentclass[doublecol,linenumbers]{epl2}
\usepackage{graphicx}
\usepackage[english]{babel}
\usepackage{latexsym,amsmath,amscd,amssymb,graphicx,amsbsy,color,xcolor,amsfonts,amsthm}

\title{Gravitational Deflection in Relativistic Newtonian Dynamics}
\shorttitle{Lensing in Relativistic Newtonian Dynamics}
\author{ Y. Friedman and J. M. Steiner}
\institute{ Jerusalem College of Technology\\Jerusalem, Israel}
\pacs{95.30.Sf}{Relativity and gravitation}
\pacs{95.10.Eg}{Orbit determination and improvement}
\pacs{98.62.Sb} {Gravitational lenses and luminous arcs}

\abstract{
 In a recent series of papers, the authors introduced a new Relativistic Newtonian Dynamics (RND) and tested its validity by the accurate prediction of the gravitational time dilation, the anomalous precession of Mercury, the periastron advance of any binary and the Shapiro time delay. This dynamics incorporates the influence of potential energy on spacetime in Newtonian dynamics and, unlike Einstein's General Relativity, treats gravity as a force without the need to curve spacetime. In this paper, this dynamics is applied to derive the gravitational deflection of both objects with non-zero mass and of massless particles passing the strong gravitating field of a massive body. Equations for the trajectory and the resulting analytical expressions for the deflection angle, in terms of the distance and velocity at the point of closest approach to the massive object, were derived in both cases.  It is shown that with a carefully defined limit, the trajectory of a massless particle is the limiting case of that of an object with non-zero mass. In the ``weak" deflection limit, the derived expression for the deflection angle of a massless particle (photon)  reproduces the experimentally tested Einstein's formula for weak gravitational lensing of a light ray, thereby providing another test for the validity of the RND.
}

\begin{document}
\maketitle

\section{Introduction}

Although Einstein's theories of relativity originate in the logical incompleteness of Newton's laws of motion, that incompleteness itself has not been understood completely, as yet, and therefore there are still global and chronic problems which still need further clarification. It is the belief of the authors that this incompleteness arises from the fact that \textit{ the classical Newton's theory does not consider the influence of energy on spacetime}.

 This suggested the development of a new  \textit{Relativistic Newtonian Dynamics} (RND) which incorporates the influence of {any} energy (gravitational or non-gravitational) on spacetime into classical Newtonian dynamics using some of the well established principles of relativity without any \textit{a priori} assumptions. Unlike in Einstein's GR, this dynamics treats gravity as a force without the need for curving the spacetime. The RND model for motion of \textit{objects with non-zero mass} in an attractive conservative force field was developed in \cite{FSMerc},\cite{Fcent}, \cite{FSHT} and further extended for \textit{massless} particles in \cite{Fshap}. All the calculations in this dynamics are performed in an inertial (lab) frame $K$ with origin at the center of the force.

The accurate prediction of gravitational time dilation, orbit precession and Shapiro time delay are considered as three tests of GR. As shown in \cite{FSMerc} for the accurate anomalous precession of Mercury, in \cite{FSHT} for the periastron advance of any binary and in \cite{Fshap} for the Shapiro time delay, the new RND passed all these tests with flying colours.

Einstein's prediction for deflection of light by massive objects (gravitational lensing), the remaining test of GR, was verified experimentally with high accuracy \cite{CW},\cite{Lebach}, \cite{Lambert} and  \cite{Shapiro}. Thus, any valid relativistic theory must also predict the lensing formula predicted by Einstein.

In this paper, the RND is applied to calculate the deflection angle of objects with non-zero mass and of massless particles passing the strong gravitating field of a massive body. For an object with non-zero mass, the expression for the deflection angle reduces to  that of a massless object as its speed approaches the maximal transverse speed (less than the speed of light $c$) at the point of closest approach to the massive body. This is not to be expected \textit{a priori} since the transition from the dynamics of an object with non-zero mass to that of a  massless particle is by no means a continuous process. This point is incorrectly overlooked in the literature by taking the mathematical limit for a non-continuous process.

 In the ``weak" deflection limit, the derived deflection angle for a massless particle (photon) yields  Einstein's formula for gravitational lensing \cite{MTW}, \cite{Rindler}, \cite{KEK}, thereby passing also the fourth remaining test of GR.

Furthermore, the derivation of the deflection for both objects with non-zero mass and massless particles using RND, is direct and produces\textit{ exact} analytical formulae for the deflection angle without any approximations \cite{Rindler} or assumptions like the wave nature of both particles and objects \cite{Unn}  quoted in the literature.

\section{ RND  for motion in a central gravitational field}

Consider a central gravitational force field with  \textit{reduced} gravitational potential (potential on an object with unit mass) $\hat{U}(r)$ vanishing at infinity, in an \textit{inertial frame } $K$ with time  and spherical space coordinates $(t,r,\varphi,\theta)$ and origin at the center of the force. Using the symmetry of the problem, assume that the motion is in the plane $\theta=\pi/2$.
Furthermore, for convenience, define the \textit{dimensionless reduced potential} as
\begin{equation}\label{uDefSchwarz}
  u(r)=-\frac{2\hat{U}(r)}{c^2}\,.
\end{equation}

As shown in \cite{Fshap}, in order to describe the influence of  potential energy on spacetime in $K$, a metric expressing this influence is introduced. Using an extension of the equivalence principle, this influence in the neighbourhood of a point $\mathbf{x}_0$ in $K$ is quantified  via the influence of the velocity and acceleration on  the escape trajectory at $\mathbf{x}_0$. Hence, this metric can be defined from $u(r)$.

By analogy to the principle of least action (a variational principle that defines the path of motion as the path with the least value of some action), the motion in RND can be viewed  as the motion along a geodesic (path with least distance) with respect to this metric.

Since the field expressed by $u(r)$ is time and $\varphi$ independent in $K$,  on any trajectory parameterized by an affine (metric dependent) parameter $\lambda$, there exist two \textit{isolating integrals of motion} $k$ and $J$ such that
\begin{equation}\label{dtdsC}
c( 1-u(r)) \frac{dt}{d\lambda}=k
\end{equation}
and
\begin{equation}\label{Jdef}
  r^2\frac{d\varphi}{d\lambda}=J,
\end{equation}
where $k$ and $cJ$ are related to the total energy and angular momentum on the trajectory, respectively.

Combining equations (\ref{dtdsC}) and (\ref{Jdef}) one obtains the affine parameter and metric free energy-angular momentum relation depending only on the spacetime coordinates  in $K$
\begin{equation}\label{Jdeffree}
 \frac{c J}{k}=\frac{r^2}{1-u(r)}\frac{d\varphi}{dt}\,.
\end{equation}

In terms of these integrals of motion, the affine parameter free RND \textit{equation for the trajectory } $r(\varphi)$ is
\begin{equation}\label{drdphi}
  \left(\frac{J}{r^2}\frac{dr}{d\varphi}\right)^2+(1-u(r))\left(\frac{J^2}{r^2}+\epsilon\right)=k^2,
\end{equation}
where $\epsilon=0$ for a massless particle and $\epsilon=1$ for an object with non-zero mass. In both cases this equation reduces to the classical Newtonian equation for the trajectory when the coefficient  $(1-u(r))$ of the second term of this equation is replaced by 1.

\section{RND  equation for the trajectory around a spherically symmetric massive body}

Consider now the motion in inverse square law gravitational field of a spherically symmetric massive body of mass $M$ (e.g. Sun) in an inertial frame $K$, defined as above. The  Newtonian reduced  gravitational potential in this case is $\hat{U}(r)=-\frac{GM}{r}$ with the dimensionless form
\begin{equation}\label{uDefM}
u(r)=\frac{2GM}{c^2r}=\frac{r_s}{r},\;\;\;r_s=\frac{2GM}{c^2}\,,
\end{equation}
where $r_s$ is the \textit{Schwarzschild radius} of the massive body.

Suppose that a massless particle or an object with non-zero mass passes the gravitational field of this massive body from an emitter at point $E$ to a receiver at point $R$.
Denote by $r_0$ the distance from the point $P$ on the trajectory closest to the massive body, (see Figure 1). For simplicity, rotate the plane of motion so that $r_0=r(\pi/2)$.

\begin{figure}[h!]
\centering
 \scalebox{0.28}{\includegraphics{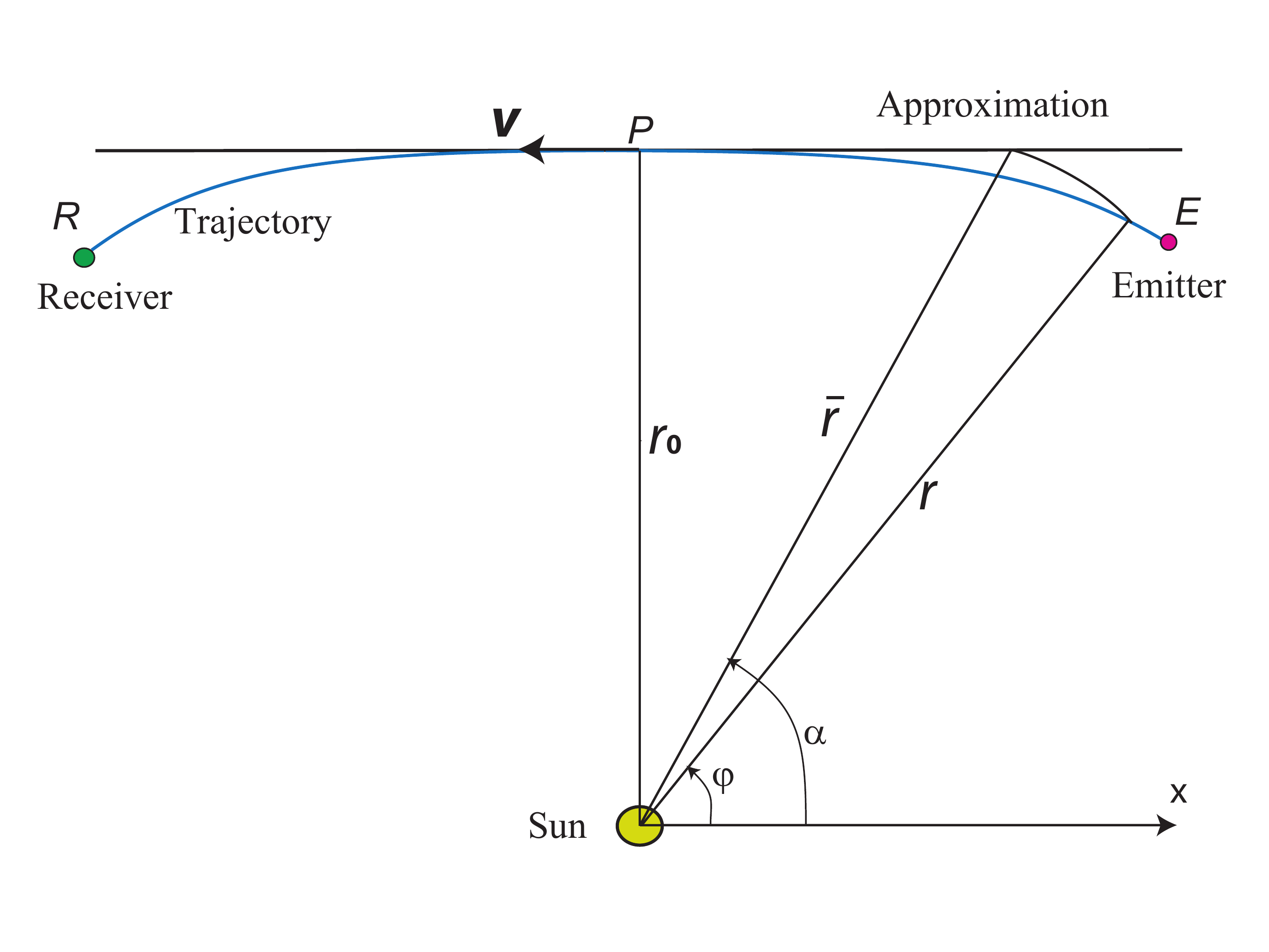}}
 \caption{The bending trajectory, the straight line approximation and associated angles $\varphi,\alpha$ }\label{Prec}
\end{figure}

\noindent\textbf{1. Massless particles ($\epsilon=0$).} In terms of $r_s$ the equation of the trajectory (\ref{drdphi}) becomes
\begin{equation}\label{drdphiPh}
   \left(\frac{J}{r^2}\frac{dr}{d\varphi}\right)^2+(1-\frac{r_s}{r})\frac{J^2}{r^2}=k^2.
\end{equation}
Since $\frac{dr}{d\varphi}=0$ at the point of closest approach $P$, one obtains
\begin{equation}\label{JPh}
  \frac{J^2}{k^2}=\frac{r_0^2}{1-\frac{r_s}{r_0}}\,.
\end{equation}

Substituting this into (\ref{drdphiPh}) yields (the integrals of motion free) \textit{equation for the trajectory of massless particles}
\begin{equation}\label{Ph2}
  \left(\frac{r_0}{r^2}\frac{dr}{d\varphi}\right)^2+(1-\frac{r_s}{r})\frac{r_0^2}{r^2}=1-\frac{r_s}{r_0}
\end{equation}
in terms of $r_0$.

\noindent\textbf{2. Objects with non-zero mass ($\epsilon=1$).}
 In terms of $r_s$, the equation of the trajectory (\ref{drdphi}) of an object with non-zero mass becomes
\begin{equation}\label{drdphiMb}
   \left(\frac{J}{r^2}\frac{dr}{d\varphi}\right)^2+(1-\frac{r_s}{r})\left(\frac{J^2}{r^2}+1\right)=k^2.
\end{equation}
At the point of closest approach $P$, $\frac{dr}{d\varphi}=0$ implying that the velocity $\mathbf{v}$ of the object at $P$, as measured in $K$,  is transverse to the radial direction and $v=r_0\frac{d\varphi}{dt}$, whence  from (\ref{Jdeffree}) one obtains
\begin{equation}\label{Jmass1}
  \frac{J}{k}=\frac{r_0}{1-r_s/r_0}\frac{v}{c}\,,
\end{equation}
which upon substitution  into (\ref{drdphiMb}) yields
\begin{equation}\label{drdphiM}
   (1-\frac{r_s}{r_0})\frac{k^2}{(1-r_s/r_0)^2}\left(\frac{v}{c}\right)^2+1-\frac{r_s}{r_0}=k^2.
\end{equation}
From this equation, the integral of motion $k$, in terms of the distance and velocity at the point of closest approach, becomes
\begin{equation}\label{kVal}
 k^2=\frac{(1-r_s/r_0)^2}{1-r_s/r_0-(v/c)^2}\,.
\end{equation}

Since $k^2>0$, this imposes a limitation on the  speed of the object at $P$
\begin{equation}\label{cp}
v< c_p,\;\;\;\; c_p =c\sqrt{1-r_s/r_0}\,,
\end{equation}
where $c_p$ denotes  \textit{the maximal transverse speed} at the point $P$ in $K$. It was shown in \cite{Fcent}, \cite{FSHT}  that   $v=v_0\sqrt{1-r_s/r_0}$, where $v_0$ is the speed of the moving object in the local frame at $P$ influenced by potential energy. Hence, the above limitation reveals that $v_0=\frac{v}{\sqrt{1-r_s/r_0}}< c$, i.e. in the local frame the speed of an object with non-zero mass is limited by the speed of light $c$.

 Since the velocity of a massless  object belong to the boundary of the domain of relativistically admissible velocities for objects with non-zero mass,  the speed of a photon moving transversely to the radial direction at $P$ is $c$ in this local frame and  $c_p<c$ in the inertial frame $K$. This observation explains the Shapiro time delay measured in the inertial frame $K$ where the speed of the light in the proximity of the massive body is less then $c$.

Note that equation (\ref{Jmass1}) relating the integrals of motion for objects of non-zero mass, reduces to its counterpart (\ref{JPh}) for a massless particle in the limit $v\to c_p$   (with $c_p$ defined in (\ref{cp}))
\begin{equation}\label{LimPh}
   \frac{J}{k}=\lim_{v\to c_p}\frac{r_0}{1-r_s/r_0}\frac{v}{c}=\frac{r_0}{\sqrt{1-r_s/r_0}}.
\end{equation}

Dividing (\ref{drdphiMb}) by $k^2$, substituting (\ref{Jmass1}), (\ref{kVal})  and multiplying  by $(c/v)^2(1-r_s/r_0)^2=(c_p/v)^2(1-r_s/r_0),$  one obtains the (integrals of motion free) \textit{equation for the trajectory of objects with non-zero mass}
\begin{equation}\label{TrajM}
 \left( \frac{r_0}{r^2} \frac{dr}{d\varphi}\right)^2+(1-\frac{r_s}{r})\left(\frac{r_0^2}{r^2}+(\frac{c_p}{v})^2-1 \right)
=(\frac{c_p}{v})^2\left(1-\frac{r_s}{r_0}\right)
\end{equation}
in terms of $r_0$ and $v$.

It is worthwhile noting that although, in terms of the integrals of motion, the trajectory equations (\ref{drdphiPh}) and (\ref{drdphiMb}) for massless particles and objects with non-zero mass are  different,  their resulting counterparts (\ref{Ph2}) and  (\ref{TrajM}) with the integrals of motion expressed in terms of  the distance and velocity at the point of closest approach  are identical in the limit $v\rightarrow c_p$. Thus, the trajectory of a massless particle can be obtained by applying this limit to the trajectory of an object with non-zero mass.

\section{Gravitational deflection by a massive body in RND}

Equation (\ref{TrajM}) can be rewritten as
\begin{equation*}
 \left( \frac{r_0}{r^2} \frac{dr}{d\varphi}\right)^2= (\frac{c_p}{v})^2\frac{r_s}{r_0}\left(\frac{r_0}{r} -1\right)+\left(1-\frac{r_s}{r}\right)\left(1-\frac{r_0^2}{r^2}\right)=
\end{equation*}
\begin{equation}\label{Mtraj}
=1-\frac{r_0^2}{r^2}-\frac{r_s}{r_0}\left(\frac{r_0}{r}-\frac{r_0^3}{r^3}+(\frac{c_p}{v})^2(1-\frac{r_0}{r}) \right).
\end{equation}

Using the method for the description of the periastron advance in binaries presented in \cite{FSHT}, for any angle $\varphi$ on the trajectory one associates an angle $\alpha(\varphi)$ for which $r(\varphi)=\bar{r}(\alpha)$ where $\bar{r}(\alpha)=\frac{r_0}{\sin \alpha}$ is the polar form of the straight line  approximation of the trajectory at the point $P$ (see Figure 1).

This defines the trajectory as $ r(\varphi)=\frac{r_0}{\sin\alpha(\varphi)},$
suggesting the natural substitution
\begin{equation}\label{subst}
   r=\frac{r_0}{\sin\alpha},
\end{equation}
which implies
\begin{equation}\label{Impsubst}
\frac{dr}{d\varphi}=-\cos\alpha \frac{r^2}{r_0}\frac{d\alpha}{d\varphi}.
\end{equation}

 Substituting these into (\ref{Mtraj}), taking the square root and dividing  by $\cos\alpha$, one obtains
\begin{equation}\label{dalfdpM}
  \frac{d\varphi}{d\alpha}=\left( 1-\frac{r_s}{r_0}\left(\sin \alpha+\frac{(\frac{c_p}{v})^2}{1+\sin\alpha}\right)\right)^{-1/2}
\end{equation}
with the positive sign chosen since $\varphi$ increases with $\alpha$.

Hence, the angular change in moving from $E$ to $R$ is
\begin{equation*}
  \phi=\int_{\alpha_E}^{\alpha_R}\left(1-\frac{r_s}{r_0}\left(\sin \alpha+\frac{(\frac{c_p}{v})^2
  }{1+\sin\alpha}\right)\right)^{-1/2}d\alpha,
\end{equation*}
where $\alpha_E, \alpha_R$ are the $\alpha$ values of the points $E$ and $R$, respectively.
Assuming that these points are very remote from the massive body and subtracting the angular change for the motion along the straight line (the path if the massive object was not present), yields finally  the exact RND analytical expression for the \textit{deflection angle of an object with non-zero mass} moving from $E$ to $R$
\begin{equation}\label{phiAlphaM}
  \delta\phi=\int_{\alpha_E}^{\alpha_R}\left(1-\frac{r_s}{r_0}\left(\sin \alpha+\frac{(\frac{c_p}{v})^2
  }{1+\sin\alpha}\right)\right)^{-1/2}d\alpha-\pi.
\end{equation}

For an unbounded trajectory the speed of the object $v$ is comparable with  $c_p$. The ``weak" deflection limit assumes $\frac{r_s}{r_0}\ll 1$  (less than $4\cdot10^{-6}$ for the Sun). For $E$ and $R$ very remote from the massive body, $\alpha_E\approx 0$ and $\alpha_R\approx \pi$.  Thus, in this limit equation (\ref{phiAlphaM}) is
 \begin{equation}\label{phiAlphaM2}
  \delta\phi=\int_0^{\pi}1+\frac{r_s}{2r_0}\left(\sin \alpha+\frac{(\frac{c_p}{v})^2
  }{1+\sin\alpha}\right)d\alpha-\pi.
\end{equation}
 Using that $\int_0^{\pi} \sin \alpha d\alpha =\int_0^{\pi}\frac{1}{1+\sin\alpha}d\alpha=2$,  the \textit{weak deflection angle of an object with non-zero mass} is
\begin{equation}\label{RNDdeflecM}
  \delta\phi \approx\frac{r_s}{ r_0}\left(1+\frac{c_p^2}{v^2}\right)=\frac{2GM}{ r_0}\left(\frac{1}{c^2}+\frac{1}{v_0^2}\right),
\end{equation}
which the known formula \cite{MTW} and \cite{Unn} for the weak deflection of an object with non-zero mass.

For massless particles the \textit{weak deflection angle} is obtained by substituting $v=c_p$ in (\ref{RNDdeflecM}), resulting in
\begin{equation}\label{RNDdeflec}
   \delta\phi\approx\frac{2r_s}{r_0}=\frac{4GM}{c^2 r_0}
\end{equation}
which is identical to Einstein's formula \cite{Rindler}, \cite{KEK} and \cite{Unn} for weak gravitational lensing using GR.

\section{Results and Discussion}

The Relativistic Newtonian Dynamics (RND) was presented by the authors in \cite{FSMerc}, \cite{Fcent} and \cite{FSHT} and further refined in \cite{Fshap}. This dynamics incorporates the influence of potential energy on spacetime in Newtonian dynamics and unlike Einstein's GR, treats gravity as a force without the need to curve spacetime. This dynamics was
 successfully validated by the accurate prediction of the gravitational time dilation, the anomalous precession of Mercury, the periastron advance of any binary and by the Shapiro time delay - three known tests of Einstein's GR.

In this paper the RND  is further validated by applying it to the problem of gravitational deflection of both massless particles (gravitational lensing) and of objects with non-zero mass, passing the strong gravitating field of a massive body - the remaining test of GR. In both cases, the equations for the trajectory in the gravitational field of a spherically symmetric massive body with Schwarzschild radius $r_s$ were first obtained in terms of the integrals of motion (\ref{drdphiPh}) and (\ref{drdphiMb}).

Then, expressing these integrals of motion in terms of $r_0$-the distance of the point of the closest approach and the velocity $v$ at this point, the trajectory equations (\ref{Ph2}) and  (\ref{TrajM}) were expressed in terms of these conditions.

We discovered that the speed of an object with non-zero mass moving transversally to the radial direction at a point $P$ is limited by the speed of light $c$ in the local frame influenced by the potential energy, and by $c_p$, defined in (\ref{cp}), which is less then $c$, in the inertial lab frame is limited. It is worthwhile noting that although, in terms of the integrals of motion, the trajectory equations for massless particles and objects with non-zero mass are  different, the integrals of motion free trajectory equation for an object with non-zero mass reduces to that of a massless  ray of light  in the limit $v \rightarrow c_p$. As a consequence, this is is also true for the deflection angle.

From  equation (\ref{TrajM}), an exact analytic expression for the deflection angle for an object with non-zero mass (\ref{phiAlphaM}) was derived. By applying the above limit, this yields the exact analytic expressions for the deflection angle for a massless particle. Finally, using these exact expressions, the corresponding formulae (\ref{RNDdeflecM}) and (\ref{RNDdeflec}) in the weak deflection limit  $\frac{r_s}{r_0}\ll 1$  were obtained using the first order approximation in  $\frac{r_s}{r_0}$. Formula (\ref{RNDdeflec}) is identical to Einstein's formula for weak lensing, which further confirms the validity of the RND.

In the paper we clarified why the trajectory of a massless particle is an appropriate limiting case of that of an object with non-zero mass. This cannot be assumed \textit{a priori} (as commonly assumed in the literature) since, as shown in this paper, the transition from the dynamics and integrals of motion for an object with non-zero mass to that of a massless particle is not a continuous process, hence taking such mathematical limit should be justified.

\end{document}